\documentclass[
10pt,twocolumn,notitlepage,oneside]{revtex4}
\usepackage{rotating}
\usepackage{amsfonts,longtable}
\usepackage{amsmath}

\def\kms{$\textrm{km/s}$}

\def\H2{H$_{2}$}

\def\roH2{$\rho_{\textrm{H}_2}$}

\def\MH2{M$_{\textrm{H}_2}$}

\begin{document}

\title{Galactic disks and their evolution}
\author{\firstname{A.~V.}~\surname{Zasov}}
\email[]{zasov@sai.msu.ru}
\affiliation{Sternberg Astronomical Institute Lomonosov Moscow
University }
\author{\firstname{O.~K.}~\surname{Sil'chenko}}
\affiliation{Sternberg Astronomical Institute Lomonosov Moscow
University}
%
\begin{abstract}
We consider the key problems related to measuring the mass of
stellar disks and dark halos in galaxies and to explaining the
observed properties of disks formed in massive dark halos.

Published in \textit{Uspekhi Fizicheskikh Nauk 180 (4) 434-439
(2010)}

\textit{DOI: 10.3367/UFNr.0180.201004i.0434}

\end{abstract}
\maketitle
%

Rotating stellar-gas disks are the main structural elements of the
majority of the observed galaxies. They contain mostly baryonic
matter: stars and the diffuse interstellar medium (in the latter,
cold gas dominates). Disks have a large angular momentum and rotate
such that the local angular velocity decreases along the radius. The
maximum rotational velocity of the disk depends on the total mass
(luminosity) of the galaxy and is usually $100-300$ \kms, which
corresponds to the orbital period $200-300$ Myr.

Disks of galaxies are inhomogeneous structures. They contain stars
with different masses and ages, and the youngest stars are located
near the equatorial plane of the disk: there, on the bottom of the
disk gravitational potential well, the interstellar gas is stored.
But because the entire disks are old structures, their spectra are
dominated by very old stars with the age above 8 Gyr. The oldest
stars form the so-called thick disk, which is two to three times
thicker than the main stellar disk, but the mass of the thick disk
is relatively small. In fact, the formation of disks in most
galaxies has not yet been completed, because star formation is
occurring (predominantly in the spiral arms) there even at present,
but the observed rates of stellar population mass growth, except for
rare cases, are very low, about 1-5 solar masses per year for an
entire galaxy like our own.

The origin of galaxies and their disk formation mechanism remain
unclear and are actively discussed in the literature. The very form
of the disks clearly suggests that they were formed as a result of
the evolution of a dissipative medium (gas), which had been losing
its energy by radiation with angular momentum conservation, and the
age of the disks suggests that they already existed in the very
early times of modern galaxies and had a very intensive star
formation rate at that time. Disk galaxies, which frequently have
big star- forming regions that appear from large distances like a
collection of individual bright spots, are indeed abundantly present
among galaxies with redshifts $z > 1$; we observe them at the time
of their youth.

In modern galaxy formation theories, it is important that star
formation in galaxies occurred in the gravitational field formed by
dark matter, or the so-called hidden mass, which presently must form
massive halos around galaxies extending far away from their visible
parts. Numerical simulations showed that the role of the dark halo
is decisive in both the disk formation and its later evolution. But
it is not an easy task to decompose the observed mass of a galaxy
into gas, stellar, and dark matter components. This problem can be
solved for individual galaxies by improving upon methods of
measuring the kinematical characteristics of the disk, on the one
hand, and models of evolution of stellar population spectra, on the
other hand.

The scientific boom that gave rise to an avalanche of studies of
hidden mass inside galaxies started when observational data on
galaxy disk rotational velocity curves $V(R)$ as a function of
radius R became available for sufficiently large distances from the
center. In the optical band, either classical diffraction
spectrographs with a long slit, or scanning Fabry-Perot
interferometers in a high order of interference are typically used.
The interferometers do not have slits, and the Doppler shifts of
spectral lines can be measured simultaneously at many thousands of
points on the galactic disk; using a complicated mathematical data
processing then allows recovering the two-dimensional radial
velocity field and obtaining the rotational curve. In Russia, such
observations have been carried out on the 6-meter telescope at the
Special Astrophysical Observatory RAS. Radio observations in the
emission lines of atomic hydrogen or molecules have a lower angular
resolution than in the optical lines, while optical observations can
be used to measure radial velocities of both gas and stars with a
high angular resolution. Nevertheless, radio velocity curves in
gas-rich galaxies are traced to much longer distances than the
optical ones, sometimes reaching far beyond the visible limits of a
galaxy, because gas disks are frequently much larger than stellar
ones. It was found that rotational velocities at large distances
from galactic centers do not decrease, as a rule, but become almost
flat (come to a plateau), or even increase with $R$.

That the rotational velocity reaches a plateau is frequently
considered if not the proof, at least a decisive argument in favor
of the existence in galaxies of a dark halo with a mass comparable
with or even exceeding that of visible matter. In fact, this is not
precisely the case because the form of the rotational velocity
curve, for some reason, does not automatically imply the presence of
dark matter. A rotational curve of any form, increasing or
decreasing, can be explained by the presence of only one disk, and
it can reflect only the peculiarity of mass distribution inside it.
We illustrate this with simple examples.

If the disk density along the radius were conserved or decreased
very slowly, the rotational velocity of the disk would increase
without bound as R increases, even in the absence of a halo. Of
course, the surface density is not constant in real galactic disks;
it quite rapidly decreases with the distance from the center. But
there is no major problem explaining the plateau on the rotational
curve. The classical example is the so-called Mestel disk. This is a
thin axially symmetric disk whose surface density $\Sigma (R)$
decreases from the center to the periphery as $1/R$. It can be shown
theoretically that the circular velocity, which in the general case
is determined by the radial gradient of the gravitational potential,

$$V^2(R) =  R\cdot \partial\Phi(R) / \partial R,$$

is independent of $R$ for the Mestel disk, and the rotational curve
of the galaxy is a horizontal line from zero to infinity, such no
dark halo is to be invoked! The mass of such a disk within any given
radius $R$ is $M(R)~=~V^2R/G$, i.e., the same as for a spherically
symmetric density distribution. In other words, it is impossible to
distinguish the Mestel disk from a spherically symmetric galaxy
using the form of the rotational velocity curve.

In that case, what indeed can be considered relevant to the presence
of a hidden mass in disk galaxies, in particular, a massive dark
halo? First, this is a discrepancy between the measured rotational
velocity curve of the galactic disk (using both the form and the
absolute value of velocities) and the expected curve calculated by
assuming that the galaxy consists of only `luminous' matter, i.e.,
of the directly observed components. Because most of the mass of the
disk resides in stars, the brightness distribution of the stellar
disk reflects its mass distribution, especially if the brightness is
measured in the near-infrared spectral band, where emission from old
stars dominates.

\begin{figure}[t]
\begin{center}
\includegraphics[width=8cm]{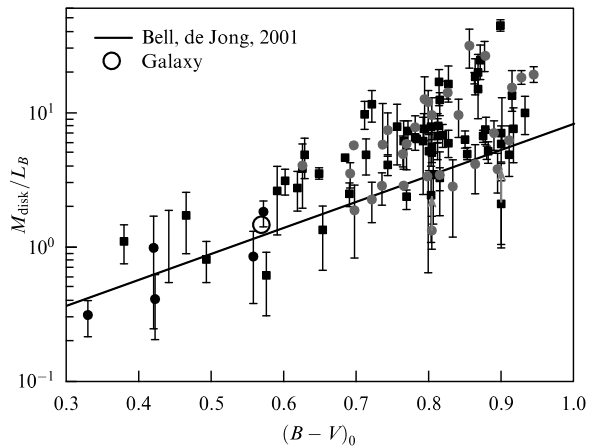}
\caption{Figure 1. The upper limit on the mass±luminosity ratio in
the blue spectral range $M/L_B$ (in solar units) for galactic disks
with different color indices, derived under the assumption of
marginal disk stability [7]. The black circles show galaxies in
pairs, the black squares mark isolated galaxies or members of
groups, and the white circle stands for our Galaxy. The values of
$M/L_B$ for galaxies with active star formation ($(B-V)_0 > 0.7$)
are close to those expected from the photometrical model of stellar
systems evolution (the straight line according to [3]), which
suggests the absence of strong dynamical heating of their disks. }
\end{center}
\end{figure}

The brightness and hence the surface density in the broad distance
range from the center decreases with $R$ exponentially as $I(R)\sim
\exp (-R/R_0)$ ($R_0$ is the radial scale of the brightness), i.e.,
more rapidly than $1/R$. This law results in a rotational curve
reaching a maximum at $R\approx 2R_0$ and then slowly decreasing.
However, the expected maximum is never observed in real galactic
rotational curves. In a more rigorous approach, by adopting a
certain mass-luminosity ratio $M/L$ for the stellar population of
the disk (this ratio can be estimated from the observed color
indices using stellar population models), the brightness
distribution can be easily transformed into the density distribution
along the disk (not necessarily exponential) and the expected
rotational velocity of the galaxy due to its baryonic components can
be calculated. In this way, the calculated rotation velocity curve
typically passes lower than the observed one (at least in the outer
parts of the disk). This allows concluding that dark mass
significantly contributes to the total mass of the galaxy. This
contribution is most substantial in galaxies with low surface
brightness of the disk: in these galaxies, dark mass can exceed the
baryonic mass by several times within the observed boundaries.

There are several other arguments in favor of high-mass dark halos
in galaxies. Two main arguments follow from observations, although
they are statistical. The first argument is that the integral masses
of galaxies and galaxy systems measured by other means (using the
relative velocities of the motion of satellites) are found to be
much larger than their visible mass. For example, the data in [1, 2]
show that the mean ratio of the total mass of galaxies in pairs to
their total infrared K-luminosity (2.2 mm), derived for more than
500 pairs, is very high: around 11 solar units, and for groups of
galaxies, two times as high. For compar- ison, models of the purely
stellar population yield the value of $M/L_K$ below 1.5 [3].
Accounting for the mass of gas and the internal extinction only
slightly increases this ratio. Consequently, in the region whose
size exceeds the diameter of the galaxy and includes a pair or group
of galaxies, the mass of dark matter exceeds the total mass of
directly visible matter by many times.

The second argument is related to the stability condition of the
disk with respect to gravitational perturbations. The disk is
usually described as a massive strongly oblate stellar system in
which oscillations can propagate and instabilities can develop on
different scales. This inevitably leads to the dynamic heating of
the disk up to a state with marginal stability under small
perturbations. The higher is the radial velocity dispersion of stars
constituting most of the disk mass and the faster it rotates and the
lower its surface density is, the more stable the disk is. The
critical (maximal) surface density at which the disk is still stable
is determined both analytically (under several simplifying
assumptions) and numerically. In the first approximation, the
critical surface density is proportional to the radial velocity
dispersion times the angular velocity of the disk at a given radius.
Hence, after having obtained the rotational velocity curve and
velocity dispersion of old stars of the disk from observations, it
is possible to estimate the maximum admissible surface density of
the disk and the corresponding $M/L$ ratio, and then to find the
upper limit of its mass from the total disk luminosity.

The accuracy of this estimate for an individual galaxy is not very
high, up to a factor of two, but data obtained for different
galaxies allow to make some general conclusions. Specifically, it
has been confirmed that within the optical limits of a galaxy, the
dark halo mass is typically comparable to that of the disk and often
exceeds it [4], which is compatible with results derived from the
analysis of rotational velocity curves. The same conclusion can be
obtained from the photometric estimates of the thickness of stellar
disks observed edge-on [5, 6]. There is another intriguing fact:
because galactic disks are subjected to gravitational perturbations
from both neighboring galaxies and massive dark halos (see below),
it follows that, apparently, gravitational perturbations must have
dynamically heated up the disk above the stability limit. Then the
$M/L$ ratios calculated under the assumption of marginally stable
disks (Fig. 1) exceed the values derived from the color index of
photometrical models of the evolution of galactic stellar disks (the
straight line in Fig. 1). It turns out that such overheated systems
do exist, but only some fraction of galaxies, predominantly with
high color indices (corrected for the disk inclination to the line
of sight) $(B-V)_0 > 0.7$ relate to them. Such a color index
corresponds to evolved disks whose luminosities are dominated by an
old stellar population. In those galaxies, the star formation is
very weak or totally absent; many of them are lenticular systems
that contain almost no cold interstellar gas. Significantly
`overheated' disks are frequently observed in galaxies in pairs
(circles in Fig. 1), obviously because they can experience a
stronger gravitational perturbation from the companion. But among
galaxies with `overheated' disks, there are galaxies without close
companions, and the increase in the velocity dispersion of their
stars can then be due to the merging of the companions that are no
longer observed as individual galaxies. But most important is that
many lenticular and most spiral galaxies had no significant
dynamical evolution over the several billion years of their lives,
and their disks are kept weakly `heated' in a state close to the
marginal stable one. The presence of galaxies with enigmatically
thin stellar disks points to the same fact. These galaxies show low
vertical velocity dispersion with respect to the rotational velocity
and are frequently found among galaxies observed edge-on (Fig. 2).

\begin{figure}
\begin{center}
\includegraphics[width=8cm]{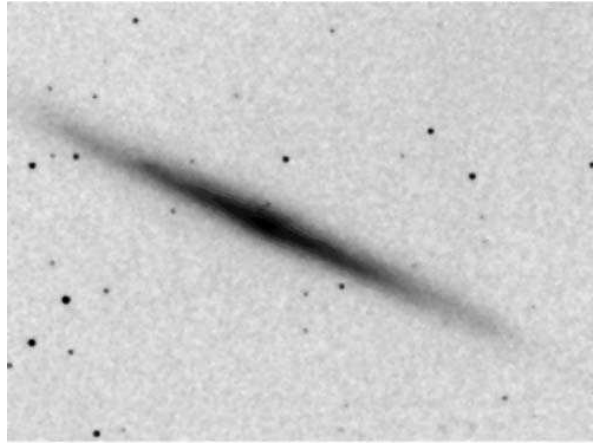}
\caption{Figure 2. NGC 5907, an example of a galaxy with a thin disk
observed edge-on (2MASS (Two Micron All-Sky Survey), 2 mkm, near
infrared). }
\end{center}
\end{figure}

Yet another argument favoring the existence of dark halos and dark
matter in the Universe in general has a rather theoretical
character: it is impossible to calculate the physical picture of
galaxy formation within the standard theory of the expanding
universe without assuming that most matter in the Universe is a
nonbaryonic dark matter. The proper gravity of baryonic matter,
which amounts to four percent of the critical density of the
Universe, is far too insufficient to explain how minuscule
primordial fluctuations could grow in a short time to form the
observed galaxies and their systems.

In the framework of the so-called standard cosmological model,
galaxies appear as a result of hierarchical clustering of numerous
dark matter units (subhalos) in gravitational fields in which the
primordial gas is concentrated, cools down, and then forms stellar
galaxies. Later on, a long evolutionary path of the galaxy begins:
the structure of the galaxy, the content and chemical abundance of
gas and stars, and star formation rate can significantly change over
several billion years. There are numerous problems from the
theoretical standpoint: the role of processes such as the
interaction of galaxies between themselves and with the
intergalactic medium, galactic mergers, the activity of galactic
nuclei, gas ejection from galaxies, and gas accretion on galactic
disks that is capable of maintaining the current star formation rate
for a long time remains unclear. In all these cases, the relative
mass of dark matter in galaxies plays either a significant or a
decisive role.

Numerical modeling of the galaxy formation process from dark matter
and baryonic matter allowed finding, at least on the qualitative
level, an explanation for the observed large-scale structure of the
Universe and distribution of galactic mass. The galactic mass-size
dependence [8] or mass concentration within the central kiloparsec,
which is much lower than that predicted by numerical models (the
so-called central cusp problem), are explained much more poorly. But
one of the most relevant present-day problems of the hierarchical
clustering model is the presence of a large number of purely disk
galaxies in the nearby Universe (see, e.g., the statistics of the
APM (Automated Plate Measuring) survey aimed at morphological
classification of galaxies from their images [9]), i.e., galaxies
without a significant central spheroid component, with only thin
stellar disks. In many spiral galaxies, including our own, these
thin stellar disks are also very \emph{old}: the age of the oldest
open clusters of our thin disk approaches 8-9 Gyr [10], and this
means that starting from the time corresponding to the redshift $z =
1$, our Galaxy has not been seriously `disturbed'.

The above proximity of the velocity dispersion of the disks of many
galaxies to the minimum value required by gravitational stability,
as well as the small velocity dispersion of disk stars in our
Galaxy, also suggest the absence of strong dynamical heating of many
galactic disks (see, e.g., [11]). This directly contradicts the
hierarchical concept that predicts permanent galactic mergings
during the evolution of the Universe. When small halos merge to
produce a $10^{12}$ solarmass halo (as in our Galaxy), the merged
fragments preserve their identity inside the big halo for a long
time. For example, recent GHALO (Galactic Halo) calculations [12]
numerically modeling a small volume about $400$ kpc in size with
high spatial resolution have identified up to a hundred thousand
`subhalos' with continuous mass spectrum inside our dark halo. Dark
matter clumps move inside the large halo on elongated orbits and
inevitably cross the baryonic galactic disk. This `bombardment'
heats up the stellar disk, which gradually thickens due to an
increase in the vertical velocity of stars. Moreover, dark matter
clumps inside which stars emerged (dwarf satellites of our Galaxy)
lose energy in encounters due to dynamical friction and finally fall
on the disk to merge with it. According to recent calculations [13],
a typical model galaxy similar to the Milky Way had to undergo about
six `minor' mergers during the last 8 Gyr (the time corresponding to
$z = 1$), one of the mergers being with a satellite whose mass is
about 10\% of the mass of our Galaxy, which would thicken its disk
by several times. In other words, thin stellar disks do not survive
on a timescale of several billion years if they are plunged into
dark matter halos as predicted by the theory; nevertheless, most
nearby galaxies do have thin disks. This contradiction between
theory and observations has not been resolved yet.

The key feature of the evolution of galactic disks is the permanent
accretion of gas from outside, `feeding' the star formation. The
need for external gas accretion at a rate roughly comparable with
that of star formation follows from many observational facts. In
particular, the scenario of chemical evolution of the disk of our
Galaxy cannot be built without considering an appreciable accretion
of gas from outside (see, e.g., [14]). The star formation rate in
the Galaxy disk over the last $9-10$ Gyr is nearly constant averaged
over $1-2$ Gyr), which indirectly points to permanent gas accretion.
Because stars synthesize all elements heavier than beryllium in the
course of their evolution, the last generation of stars must be much
more metal-abundant than $8-10$ Gyr old. However, no significant
anticorrelation of of metallicity with age is observed [15].
Moreover, there is the so-called G-dwarf problem: in a galaxy disk,
stars of spectral class G with a mass around one solar mass or
slightly lighter, among which stars of all ages are present, have
almost the same metallicity within measurement errors [16]. It seems
that the chemical evolution in the disk of our Galaxy over the last
$8-10$ Gyr `did not go forth,' although nuclear reactions in the
stars have undoubtedly occurred. This problem is resolved by
introducing accretion from outside, i.e., by assuming that gas with
minimal (and even better, zero) metal abundance has fallen onto the
disk: such a low-metallicity gas `dilutes' the gas enriched by
synthesized heavy elements and sustains the mean metallicity of the
interstellar medium at a roughly constant level.

In spiral galaxy disks, including our own, there is a notable
`metallicity gradient': the mean heavy-element abundance in both
stars and gas is higher at the center of the disk and decreases
toward the disk periphery. Qualitatively, this is clear: in central
parts of galaxies, star formation has already exhausted all the gas,
i.e., was very effective, and on the periphery, a large amount of
fresh gas remains, i.e., star formation has occurred very slowly.
Naturally, star formation is then more effective at the disk center
and more weak on its periphery. A model of the chemical evolution of
the disk of our Galaxy with variable gas accretion along the radius
was constructed in [17]. It was concluded there that the
characteristic time of accretion in which the local disk density
increases significantly (by $e$ times) linearly increases along the
radius: it is below 2 Gyr at the center, about 8 Gyr near the Sun,
and significantly exceeds the Hubble time on the far periphery (the
disk only starts forming there). The characteristic times of times
of large-scale star formation change correspondingly. This concept
was dubbed `inside-out,' meaning that the (disk) galaxy was formed
inside out. This concept has been confirmed by many observational
facts.

One of the most spectacular confirmations was obtained by the
ultraviolet space telescope GALEX (Galaxy Evolution Explorer), which
obtained images of a large sample of nearby galaxies with good
sensitivity and reasonable space resolution in the far ($\lambda
_{eff} =1516$\AA) and near ($\lambda _{eff} =2267$\AA) ultraviolet
spectrum [18]. It turned out that many disk galaxies have much
larger sizes in the ultraviolet band than in the visible range [19].
What does this mean? Young massive stars are the main `contributors
of ultraviolet' in galaxies: they have high temperatures well
exceeding $10,000$ K, and hence the energy is mainly released in the
ultraviolet range. After the completion of the galaxy survey by
GALEX, it turned out that star formation occurs in external regions
of galaxy disks, where there are almost no old stellar populations
and nothing is visible in the optical range. This means that the
disks are indeed built up, or more precisely, built over in the
outer parts, in front of our very eyes.

Recently, an interesting study was carried out to test this scenario
[20]. The evolution of ultraviolet (i.e., star-forming) disks was
observed by directly comparing the sizes of galaxies at different
redshifts. How can this be done? Because of a finite speed of light,
the more distant the galaxy is, the earlier epoch is observed. For
example, a galaxy at the redshift $z = 0.5$ is seen five Gyr while
the time delay for $z = 1$ is eight Gyr. This means that using
modern large-aperture telescopes, we can directly probe two thirds
of the age of the Universe and can observe the evolution of galaxies
galaxies during most of their lives. Due to the redshift, the proper
ultraviolet emission from a remote galaxy can be observed in the
optical range. The authors of [20] inspected the change in the
characteristic shape of the disk surface brightness radial
distribution with $z$ by adjusting the observed wavelength to the
redshift such that in the comoving frame of the galaxy, the
measurements each time related to the same (ultraviolet) spectral
range. They started from the GALEX galaxy survey carried out at
$z\approx0$. It turned out that indeed there is an evolution: at
large redshifts, ultraviolet disks were observed to be more compact.
At earlier times, star formation proceeded at the centers of the
disks; has it now moved toward the periphery? This is exactly the
evolution that is predicted by the inside-out scenario.

However, it turned out not to be so easy: when not only scales but
also absolute levels of the ultraviolet surface brightness were
compared, it was found that the peripheries of the disks appear
almost identically at $z = 1$ and $z = 0$. The brightness profiles
at $z = 1$ are more compact or, in other words, have larger slopes
due to star formation in their centers at $z = 1$ occurring more
intensively than at $z = 0$, while no evolution of the star
formation rate at the disk peripheries is observed. That is, the
disks have not `grown' in the last eight Gyr — they have simply been
completing the star formation at the center and continued it at the
periphery. The inside-out scenario clearly needed to be improved.

Another problem of the disk formation theory is that no real gas
reservoirs for accretion have been discovered to date, although it
is difficult to doubt the reality of gas accretion on galactic
disks. In addition, it is important that the external gas also has a
low metal abundance. At some period, it was thought that during the
collapse of a dark matter halo, the primordial gas gravitationally
bound to it gradually heats up during virialization and is preserved
for a long time as a hot X-ray halo around the galaxy. By gradually
cooling down, the hot X-ray halo could provide a long steady
accretion of the primordial gas onto the entire galactic disk. Such
hot gas halos are observed in galaxy clusters, but none has been
found so far around a nearby spiral galaxy (with the possible
exception of massive bulge regions in early-type galaxies).
Moreover, the detailed gasdynamic model showed that even if such
halos exist, the known mechanisms of thermal instability of hot
virialized (i.e., equilibrated) gas are unable to provide the
required amount of cold gas clouds near galactic disks and the
steady accretion over billions of years needed to build up a
large-scale stellar disk [21].

For a long time, the outer gas source was `nominated' to be
high-velocity clouds of neutral hydrogen, which are actually
observed outside the disk of our Galaxy. However, first, their
number is too small to provide required accretion rates (at best
they give 0.1-0.2 solar masses per year, which is one order of
magnitude smaller than required to sustain the modern star formation
rate). Second, when the heavy- element abundance was estimated by
absorption lines formed in clouds serendipitously located along the
line of sight, the chemical composition of the gas of high-velocity
clouds was close to the solar one, and hence this was not the
primordial gas. Presently, most high-velocity clouds of neutral
hydrogen are thought to consist of gas ejected from the Galaxy by
so-called galactic fountains — gas outflows from active star
formation regions, in which the gas is heated up by both stellar
wind from massive stars and supernova explosions. This gas then
cools to form clouds. However, first, this is not an `addition' to
the disk, but is originally a proper part of the disk, and second,
the chemical composition is not the primordial one but is instead
enriched by the products of nucleosynthesis.

In recent years, important changes have been occurring in the theory
of galaxy formation. Hot virialized gas halos of young galaxies are
now `disfavored' as sources of matter for stellar disk formation;
theoreticians doubt that the gas virialization occurs in most of the
collapsing halos. The formation of disks and bulges of galaxies does
not necessarily have to occur via the merging of small-size
subsystems only. A more and more important role in galaxy formation
is probably played by cold filamentary gas flows directed to the
inner part of a halo [22]. This is also a sort of flow accretion,
but the accretion via gas streams that cannot occur on the entire
disk and rather fuel its periphery. These cold streams pass without
stopping through a hot gas halo and fall onto the disk. According to
modern models [23], cold flows must dominate in low-mass (relative
to the dark mass of clusters and groups of galaxies) halos at all
redshifts starting from $z = 5-6$. This means that there has been no
effective gas accretion from outside onto the \emph{center} of the
disk at any stage of galactic evolution. Therefore, the inside-out
galaxy building scenario in its classical formulation now conflicts
with both observations and the cosmological theory. Clearly, the
time for its cardinal revision is coming.

Our studies of galaxy disks are partially supported by the RFBR
grants 07-02-00792 and 07-02-00229.


\begin{thebibliography}{99}
\bibitem{1}
Makarov D, Karachentsev I Proc. Int. Astron. Union 3 370 (2008);
arXiv:0801.0043;

\bibitem{2}Karachentsev I D, Makarov D I Astrofiz. Byull.
63 320 (2008) [Astrophys. Bull. 63 299 (2008)]; arXiv:0812.0689

\bibitem{3}Bell E F, de Jong R S Astrophys. J. 550 212 (2001)

\bibitem{4}Zasov A V,
Khoperskov A V, Tyurina N V Pis'ma Astron. Zh. 30 653 (2004)
[Astron. Lett. 30 593 (2004)]

\bibitem{5}Zasov A V et al. Pis'ma Astron. Zh.
28 599 (2002) [Astron. Lett. 28 527 (2002)]

\bibitem{6}Bizyaev D, Mitronova
S Astrophys. J. 702 1567 (2009)

\bibitem{7}Zasov A V, Khoperskov A V,
Saburova A S, in preparation

\bibitem{8}Dutton A A et al. Astrophys. J. 654
27 (2007)

\bibitem{9}Naim A et al. Mon. Not. R. Astron. Soc. 274 1107 (1995)


\bibitem{10}Paunzen E, Netopil M Mon. Not. R. Astron. Soc. 371 1641 (2006)


\bibitem{11}Wyse R F G Proc. Int. Astron. Union 4 179 (2009);
arXiv:0809.4516

\bibitem{12}Zemp M Mod. Phys. Lett. A 24 2291 (2009)

\bibitem{13}
Kazantzidis Set al. Astrophys. J. 700 1896 (2009)

\bibitem{14}Tosi M,
astro-ph/0308463

\bibitem{15}Feltzing S, Holmberg J, Hurley J R Astron.
Astrophys. 377 911 (2001)

\bibitem{16}Jurgensen B R Astron. Astrophys. 363
947 (2000)

\bibitem{17}Chiappini C, Matteucci F, Gratton R Astrophys. J. 477
765 (1997)

\bibitem{18}Gil de Paz A et al. Astrophys. J. Suppl. 173 185
(2007)

\bibitem{19}Thilker D A et al. Astrophys. J. Suppl. 173 538 (2007)


\bibitem{20}Azzollini R, Beckman J E, Trujillo I Astron. Astrophys. 501 119
(2009)

\bibitem{21}Binney J, Nipoti C, Fraternali F Mon. Not. R. Astron.
Soc. 397 1804 (2009)

\bibitem{22}Dekel A, Birnboim Y Mon. Not. R. Astron.
Soc. 368 2 (2006)

\bibitem{23}Dekel A et al. Nature 457 451 (2009)

\end{thebibliography}
\end{document}